\documentclass[aps,prl,twocolumn,superscriptaddress]{revtex4}

\usepackage{graphicx}

\newcommand{\figwidth}{3.375in} 

\begin{document}
\title{Ordering of the Heisenberg Spin Glass 
in Two Dimensions
}
\author{Hikaru Kawamura and Hitoshi Yonehara}
\affiliation{Department of Earth and Space Science, Faculty of Science,
Osaka University, Toyonaka 560-0043,
Japan}
\date{\today}
\def\gsim{\buildrel {\textstyle >}\over {_\sim}}
\def\lsim{\buildrel {\textstyle <}\over {_\sim}}
\begin{abstract}
The spin and the chirality
orderings of the Heisenberg spin glass in two dimensions
with the nearest-neighbor
Gaussian coupling are investigated
by equilibrium Monte Carlo simulations.
Particular attention is paid to the behavior of the spin and the chirality
correlation lengths.
In order to observe the true
asymptotic behavior, fairly large system size $L\gsim 20$ ($L$ the linear
dimension of the system) appears to be necessary.
It is found that both the spin and the chirality order
only at zero temperature. At high temperatures, the chiral correlation length 
stays shorter than spin correlation length, whereas 
at lower temperatures below the crossover temperature
$T_\times$, the chiral correlation length exceeds the spin correlation
length. The spin and the chirality
correlation-length exponents are estimated above $T_\times$ 
to be $\nu_{{\rm SG}}=0.9\pm 0.2$ and
$\nu_{{\rm CG}}=2.1\pm 0.3$, respectively. These values are 
close to the previous estimates on the basis of the
domain-wall-energy calculation. Discussion is given about the asymptotic
critical behavior realized below $T_\times$.
\end{abstract}

\maketitle


%
%
%
%
\section{Introduction}

Spin glasses (SGs)
are the type of random magnets possessing both the ferromagnetic
and the antiferromagnetic couplings, and are
characterized both by frustration
and randomness. Experimentally, it is now well established that
typical SG magnets exhibit an equilibrium phase transition at a finite
temperature and that
there exists a thermodynamic SG phase. The true nature of the SG
transition and of the SG ordered phase, however, has not fully been
understood yet in spite of extensive studies for years.

In numerical studies of SGs, much effort has
been devoted to clarify the properties of the so-called
Edwards-Anderson (EA) model~\cite{SGrev}.
Most of these numerical works on the EA model have concentrated
on the {\it Ising\/} EA model. By contrast, many of real
SG magnets are Heisenberg-like rather than Ising-like in the sense that
the magnetic anisotropy is considerably weaker than the isotropic
exchange interaction~\cite{SGrev,OYS}. Presumably,
theoretical bias toward the
Ising model is partly due to the simplicity of the Ising model,
but partly also reflects the historical situation
that the earlier theoretical studies on the
Heisenberg EA model strongly suggested in common
that the Heisenberg EA model did not
exhibit any finite-temperature 
transition~\cite{OYS,Banavar,McMillan,Matsubara1,Yoshino}, 
apparently making it
difficult in explaining the experimental SG transition within the
isotropic Heisenberg model.

Meanwhile, a novel
possibility was suggested by one of the present authors (H.K.)
that the 3D Heisenberg SG might exhibit an equilibrium phase
transition at a finite temperature,
not in the spin sector as usually envisaged,
but in the {\it chirality\/} sector,
{\it i.e.\/}, might exhibit a {\it chiral-glass\/} (CG)
transition~\cite{Kawamura92}.
Chirality is a multispin variable
representing the sense or the handedness of local
noncoplanar spin structures induced by spin frustration.
In the CG ordered
state, chiralities are ordered in a spatially random manner while
Heisenberg spins remain paramagnetic. 
Refs.~\cite{Kawamura92,Kawamura95,Kawamura98,HK1,3DHSGinH_1,3DHSGinH_2} 
claimed that the standard SG order
associated with the freezing of the Heisenberg spin occurred at a temperature
lower than the CG transition temperature at $T=T_{\rm SG}<
T_{\rm CG}$, quite possibly $T_{\rm SG}=0$.
It means that the spin and the
chirality are decoupled on long length scales
(spin-chirality decoupling). In fact,
based on such a spin-chirality decoupling
picture, a chirality scenario of the SG transition
has been advanced,
which  explains the experimentally observed SG transition as essentially
chirality driven~\cite{Kawamura92,Kawamura98}.
Note that the numerical observation
of a finite-temperature CG transition in the 3D Heisenberg SG 
\cite{Kawamura92,Kawamura95,Kawamura98,HK1,3DHSGinH_1,3DHSGinH_2}
is not inconsistent with the earlier observations
of the absence of the conventional SG order at any finite temperature.

Recently, however, in a series of numerical studies on the 3D
Heisenberg EA model, Tohoku group criticized the earlier numerical works,
claiming that in the 3D Heisenberg SG the spin ordered at a finite temperature
and that the SG transition temperature
might coincide with the CG transition temperature,
{\it i.e.\/}, $T_{{\rm SG}}=T_{{\rm CG}}>0$~\cite{Matsubara2,Nakamura}.
By calculating the spin and the chirality correlation lengths, Lee and Young
also suggested $T_{{\rm SG}}=T_{{\rm CG}}>0$~\cite{Young}.
By contrast, Hukushima and Kawamura maintained that
in 3D the spin and the
chirality were decoupled on sufficiently long length scales, and that
$T_{{\rm SG}}<T_{{\rm CG}}$~\cite{HK2}, supporting the earlier numerical
results. The situation in 3D thus remains controversial.

Under such circumstances,
in order to shed further light on the nature of the
ordering in 3D,
it might be useful to study the
problem for the general space dimensionality $D$. The spin
and the chirality orderings of the Heisengerg SG in dimensions higher
than three, {\it i.e.\/}, 4D, 5D and $D=\infty$
(an infinite-ranged mean-field SG model),
were recently studied in Ref.\cite{ImaKawa}.
In the present paper, we wish to study the spin
and the chirality orderings of the low-dimensional system, {\it i.e.\/},
the Heisengerg SG in 2D.

There were only a few thoeretical and numerical works
performed for the 2D Heisenberg EA model. On analytical side, there
exists a proof that the standard
SG long-range order (LRO) does not arise at any finite temperature
\cite{Schwartz}.
The proof, however,  does not tell whether the CG
LRO could exist or not in 2D. On numerical side, the domain-wall-energy
calculation suggested
that both the spin and the chirality
ordered only at $T=0$, but with mutually different correlation-length
exponents, $\nu_{{\rm CG}}\sim 2.1>\nu_{{\rm SG}}\sim 1.0-1.2$
\cite{Kawamura92}. 
If this is
really the case, the spin and the chirality are decoupled on long
length scale at the
$T=0$ transition. We note that a
similar phenomenon has also been reported
for the 2D {\it XY\/} SG \cite{KawaTane,Moore,Bokil,Wengel}. Meanwhile, to the
authors' knowledge, no Monte Carlo (MC) simulation  for the 2D Heisenberg SG
has been reported so far.

In the present paper,
we wish to fill this gap. We study both the
SG and the CG orderings of the 2D Heisenberg EA model
by means of a large-scale equilibrium MC simulation.
The present paper is organized as follows.
In \S 2, we introduce our model and explain some of the
details of the MC calculation. Various 
physcial quantities calculated in our MC simulations,
{\it e.g.\/}, the SG and the CG susceptibilities,
the spin and the chiral Binder ratios, the SG and the CG
correlation functions and the associated correlation legnths are introduced in
\S 3.
The results of our MC simulations
are presented and analyzed in \S 4.
Section 5 is devoted to brief summary of the results.

%
%
\section{The model and the method}
\label{secModel}

The model we consider is the isotropic classical Heisenberg
model on a 2D square lattice
with the nearest-neighbor Gaussian coupling.
The Hamiltonian is given by
\begin{equation}
{\cal H}=-\sum_{<ij>}J_{ij}\vec{S}_i\cdot \vec{S}_j\ \ ,
\label{eqn:hamil}
\end{equation}
where $\vec{S}_i=(S_i^x,S_i^y,S_i^z)$ is a three-component unit vector,
and the $<ij>$ sum is taken over nearest-neighbor pairs on the lattice.
The nearest-neighbor coupling $J_{ij}$ is assumed to
obey the Gaussian distribution with zero mean and variance $J^2$.
We perform equilibrium MC simulations on the model.
The lattices studied are square lattices with $N=L^{2}$ sites
with $L=10$, 16, 20, 30 and 40 with
periodic boundary conditions in all directions.
Sample average is taken over 384 ($L=10,16,20$) and 320 ($L=30,40$)
independent bond realizations.
Error bars of
physical quantities are estimated by the sample-to-sample statistical
fluctuation over the bond realizations.

In order to facilitate efficient thermalization, we combine the standard
heat-bath method with the temperature-exchange technique~\cite{TempExMC}.
Care is taken to be sure that
the system is fully equilibrated.
Equilibration is checked by the following procedures.
First, we monitor the system to travel back and forth
many times along the
temperature axis during
the temperature-exchange process (typically more than 10 times)
between the maximum and minimum temperature points.
We check at the same time
that the relaxation
due to the standard heat-bath updating
is reasonably fast at the highest temperature,
whose relaxation time is of order $10^2$ Monte Carlo steps
per spin (MCS). This guarantees that different parts of
the phase space are sampled in each ``cycle'' of the temperature-exchange
run. Second, we check
the stability of the results against at least three times longer runs
for a subset of samples.
%

%
%
\section{Physical Quantities}

In this section, we
define various physical quantities calculated in our
simulations below.
By considering two independent systems (``replicas'') described by
the same Hamiltonian (\ref{eqn:hamil}),
one can define an overlap variable.
The overlap of the Heisenberg spin
is defined
as a {\it tensor\/} variable $q_{\alpha\beta}$
between the $\alpha$ and $\beta$
components ($\alpha$, $\beta$=$x,y,z$) of the Heisenberg spin,
\begin{equation}
q_{\alpha\beta}=\frac{1}{N}\sum_{i=1}^N S_{i\alpha}^{(1)}S_{i\beta}^{(2)}\ \ ,
\ \ (\alpha,\beta=x,y,z)\ \ ,
\end{equation}
where $\vec{S}_i^{(1)}$ and $\vec{S}_i^{(2)}$ are the $i$-th
Heisenberg spins of the replicas 1 and 2, respectively.
In our simulations, we prepare the two replicas 1 and 2 by
running two independent sequences of systems
in parallel with different spin initial conditions and
different sequences of random numbers.
In terms of these tensor overlaps, the SG order parameter is defined by
\begin{equation}
q_{\rm s}^{(2)} = [\langle q_{\rm s}^2\rangle]\ \ ,
\ \ \ \ q_{\rm s}^2 = \sum_{\alpha,\beta=x,y,z}q_{\alpha\beta}^2\ \ ,
\end{equation}
and the SG susceptibility by
\begin{equation}
\chi_{\rm SG}^{(2)} = Nq_{\rm s}^{(2)}\ \ ,
\end{equation}
where $\langle\cdots\rangle$ represents the thermal average and
[$\cdots$] the average over the bond disorder.
The associated spin Binder
ratio is defined by
\begin{equation}
g_{\rm SG} = \frac{1}{2}
\left(11 - 9\frac{[\langle q_{\rm s}^4\rangle]}
{[\langle q_{\rm s}^2\rangle]^2}\right)\ \ .
\label{eqn:gs_def}
\end{equation}
Note that $g_{\rm SG}$ is
normalized  so that, in the thermodynamic limit,
it vanishes in the high-temperature phase and gives unity in the
nondegenerate ordered state.

The local chirality at the $i$-th site and in the $\mu$-th
direction $\chi_{i\mu}$ may be defined for three neighboring Heisenberg spins
by the scalar,
\begin{equation}
\chi_{i\mu}=
\vec{S}_{i+{\hat{e}}_{\mu}}\cdot
(\vec{S}_i\times\vec{S}_{i-{\hat{e}}_{\mu}})\ \ ,
\end{equation}
where ${\hat{e}}_{\mu}\ (\mu=x,y)$ denotes
a unit vector along the $\mu$-th axis.
There are in total $2N$ local chiral variables.
We define the mean local amplitude of the chirality, $\bar \chi$, by
\begin{equation}
\bar{\chi}=\sqrt{\frac{1}{2N}\sum_{i=1}^N
\sum_{\mu=x,y}[\langle\chi_{i\mu}^2\rangle]}\ \ .
\end{equation}
Note that the magnitude of $\bar{\chi}$
tells us the extent of the noncoplanarity of the local spin
structures. In particular,
$\bar{\chi}$ vanishes for any coplanar spin configuration.

As in the case of the Heisenberg spin,
one can define an overlap of the chiral variable
by considering the two replicas by
\begin{equation}
q_{\chi}=
\frac{1}{2N}\sum_{i=1}^N\sum_{\mu=x,y}
\chi_{i\mu}^{(1)}\chi_{i\mu}^{(2)}\ \ ,
\end{equation}
where $\chi_{i\mu}^{(1)}$ and $\chi_{i\mu}^{(2)}$ represent the chiral
variables of the replicas 1 and 2, respectively.
In terms of this chiral overlap $q_{\chi}$, the CG
order parameter is defined by
\begin{equation}
q_{\chi}^{(2)}=[\langle q_{\chi}^2\rangle]\ \ ,
\end{equation}
and the associated CG susceptibility by
\begin{equation}
\chi_{{\rm CG}}=2N[\langle q_{\chi}^2\rangle]\ \ .
\end{equation}
Unlike the spin variable, the local magnitude
of the chirality is not normalized to unity, and is also
temperature dependent somewhat. 
In order to take account of this effect, and also to get an appropriate
normalization,
we also consider the reduced CG susceptibility
$\tilde \chi_{{\rm CG}}$
by dividing $\chi_{{\rm CG}}$ by appropriate powers of $\bar \chi$,
\begin{equation}
\tilde \chi_{{\rm CG}}=\frac{\chi_{{\rm CG}}}{\bar \chi^4}\ \ .
\end{equation}
The Binder ratio of the chirality is defined by
\begin{equation}
g_{{\rm CG}}=
\frac{1}{2}
\left(3-\frac{[\langle q_{\chi}^4\rangle]}
{[\langle q_{\chi}^2\rangle]^2}\right)\ \ .
\end{equation}
The two-point SG correlation function is defined by
\begin{eqnarray}
C_{{\rm SG}}(\vec r) & = &
\frac{1}{N}\sum_i[\langle \vec S_i\cdot
\vec S_{i+\vec r}\rangle^2], \nonumber \\ 
& = & \frac{1}{N}\sum_i\sum_{\alpha \beta}
[\langle S_{i,\alpha}^{(1)}S_{i,\beta}^{(2)}
S_{i+\vec r,\alpha}^{(1)}S_{i+\vec r,\beta}^{(2)}\rangle],
\end{eqnarray}
where $\vec r=(x,y)$ denotes the two-dimensional positional vector
between the two Heisenberg spins.
The associated spin correlation length $\xi_{{\rm SG}}$ is defined,
with $\hat C(\vec k)=\hat C(k_x,k_y)$ being the Fourier transform of 
$C(\vec r)$, by
\begin{equation}
\xi_{{\rm SG}}=\frac{1}{2\sin(\frac{\pi}{L})}
\sqrt{\frac{\hat C_0}{\hat C_m}-1},
\end{equation}
\begin{equation}
\hat C_0=\hat C(0,0),\ \ \ 
\hat C_m=\frac{1}{2}[\hat C(k_{min},0)
+ \hat C(0,k_{min})],\ \ 
\end{equation}%
where $k_{min}=\frac{2\pi}{L}$ 
is the minimum nonzero wavevector under the periodic
boundary conditions.

Concering the chirality correlation, we introduce the two-point
CG correlation function $C_{{\rm CG}}^{\mu\nu}(\vec r)$
between the two local chiral variables
in the $\mu$-th  and  in the $\nu$-th directions, which are
apart by the positional vecotr $\vec r$, by
\begin{eqnarray}
C_{{\rm CG}}^{\mu\nu}(\vec r) & = &
\frac{1}{N}\sum_i[\langle \chi_{i,\mu}
\chi_{i+\vec r,\nu}\rangle^2], \nonumber \\
& = & \frac{1}{N}\sum_i[\langle \chi_{i,\mu}^{(1)}\chi_{i,\mu}^{(2)}
\chi_{i+\vec r,\nu}^{(1)}\chi_{i+\vec r,\nu}^{(2)}\rangle].
\end{eqnarray}
We then define the following three types of chiral correlation lengths,
$\xi_{{\rm CG}}^\perp$, $\xi_{{\rm CG}}^\parallel$ and $\xi_{{\rm CG}}^+$,
depending on the relative directions of the local chiral variables
with respect to the positional vector $\vec r$,
\begin{equation}
\xi_{{\rm CG}}^\perp = \frac{1}{2\sin(\frac{\pi}{L})}
\sqrt{\frac{\hat C_{\chi 0}^\perp}
{\hat C_{\chi m}^\perp} - 1},\ \ 
\end{equation}
\begin{equation}
\xi_{{\rm CG}}^\parallel = \frac{1}{2\sin(\frac{\pi}{L})}
\sqrt{\frac{\hat C_{\chi 0}
^\parallel}
{\hat C_{\chi m}^\parallel} - 1},\ \ 
\end{equation}
\begin{equation}
\xi_{{\rm CG}}^+ = \frac{1}{2\sin(\frac{\pi}{L})}
\sqrt{\frac{\hat C_{\chi 0}^+}
{\hat C_{\chi m}^+} - 1},\ \ 
\end{equation}
where the $k=0$ Fourier components $C_{\chi,0}^\perp$, $C_{\chi,0}^\parallel$
and $C_{\chi,0}^+$, 
are given by
\begin{equation}
\hat C_{\chi0}^\perp = \hat C_{\chi0}^\parallel = 
\frac{1}{2}[\hat C^{xx}(0,0)
+\hat C^{yy}(0,0)],\ \ 
\end{equation}
\begin{equation}
\hat C_{\chi0}^+ = \frac{1}{2}[\hat C^{xy}(0,0)
+ \hat C^{yx}(0,0)],\ \ 
\end{equation}
while the $k=k_{min}$ components $C_{\chi m}^\perp$, 
$C_{\chi m}^\parallel$ and $C_{\chi m}^+$ are given by 
\begin{equation}
\hat C_{\chi m}^\perp = \frac{1}{2}[\hat C^{yy}(k_{min},0)
+\hat C^{xx}(0,k_{min})],\ \ 
\end{equation}
\begin{equation}
\hat C_{\chi m}^\parallel = \frac{1}{2}[\hat C^{xx}(k_{min},0)
+\hat C^{yy}(0,k_{min})],\ \ 
\end{equation}
\begin{eqnarray}
\hat C_{\chi m}^+ = \frac{1}{4}[\hat C^{xy}(k_{min},0)
+\hat C^{yx}(k_{min},0) \nonumber \\ 
+\hat C^{xy}(0,k_{min}) + \hat C^{yx}(0,k_{min})].
\end{eqnarray}
In these configurations, the directions of the two chiralities are, (i)
both perpendicular to the positional vector for $\xi_{{\rm CG}}^\perp$, (ii)
both parallel with the positional vector for $\xi_{{\rm CG}}^\parallel$, 
and (iii) one perpendicular to, 
and one parallel with the positional vector for 
$\xi_{{\rm CG}}^+$.

%
%
\section{Monte Carlo Results}

In this section, we present our MC results.
In Fig.1, we show the temperature dependence of (a) the
SG susceptibility $\chi_{{\rm SG}}$ , and of (b) the reduced CG
suceptibility $\tilde \chi_{{\rm CG}}$ (b), on a log-log plot.
Recall that the SG transition temperature of the present model
is expected to be at $T=0$. As can be seen from Fig.1(a),
our data of $\chi_{{\rm SG}}$
is consistent with a power-law
divergence at zero temperature, $\chi_{{\rm SG}}\approx 
T^{-\gamma_{{\rm SG}}}$,
with an exponent $\gamma_{{\rm SG}}\simeq 1.5-1.8$. 
In Fig.1(a), a straight line with a slope equal to $-1.8$ is drawn, 
which is the best $\gamma_{{\rm SG}}$ value determined from the other
physical quantities analyzed below.
We note that
the size dependence of  $\chi_{{\rm SG}}$
is normal in the sense that
$\chi_{{\rm SG}}$ gets larger as the system size $L$
is increased, and that this
tendency is enhanced at temperatures closer to $T_{{\rm SG}}=0$.

By contrast, the size dependence of the CG suceptibility
is somewhat unusual: In the temperature range $T/J\gsim0.025$,
$\chi_{{\rm CG}}$ gets smaller as $L$ is increased, 
contrary to the tendency normally expected for a deverging
quantity in the critical region.
At lower temperatures $T/J\lsim0.025$,  the
size dpendence of $\tilde \chi_{{\rm CG}}$ is changed into the normal one,
{\it i.e.\/}, it gets larger as $L$ is increased. Although
there is no proof that the CG transition temperature
of the model is $T_{{\rm CG}}=0$, if one fits the data to a power-law
divergence of the form, $\chi_{{\rm CG}}\approx
T^{-\gamma_{{\rm CG}}}$, assuming $T_{{\rm CG}}=0$,
one gets $\gamma _{{\rm CG}}\simeq 4.0$: See Fig.1(b) 
The CG susceptibility exponent
$\gamma _{{\rm CG}}$ is then more than twice the SG susceptibility exponent
$\gamma _{{\rm SG}}\simeq 1.8$.

For any $T=0$ transition with a nondegenerate ground state, 
the critical-point-decay exponent $\eta$ should be 
given by $\eta=2-d$ ($d$ the
space dimensionality), which leads to
the scaling
relation $\gamma =d\nu =2\nu $ (for $d=2$) 
between the susceptibility
exponent $\gamma$ and the associated correlation-length exponent $\nu$.
Hence,  mutually different
susceptibility exponents for the spin and for the chirality
necessarily mean mutually
different correlation-length exponents for the spin and for the
chirality, {\it i.e.\/}, $\nu_{{\rm SG}}\simeq 0.9$ and $\nu_{{\rm CG}}\simeq 
2.0$. This apparently
indicates the spin-chirality decoupling. However, one should be careful
that the present data
of the SG and CG susceptibilities
alone are not enough to completely exclude the possibility 
that the critical behavior crosses over 
to a different one at still lower temperatures, or even the possibility
of the occurrence of a finite-temperature CG transition.

In order get further information, we show in Fig. 2 (a) the Binder ratio of
the spin $g_{{\rm SG}}$, and (b) of the chirality $g_{{\rm CG}}$.
As can be seen from Fig.2(a),
the spin Binder ratio for smaller sizes $L\leq 20$
exhibits a crossing at a nonezero temperature, while
the crossing temperatures tend to shift to lower temperatures
with increasing $L$.
For larger
sizes $L\geq 20$, by contrast, $g_{{\rm SG}}$ does not exhibit a crossing
at any temperature studied,
always getting smaller with increasing $L$. Such a behavior
is indeed consistent with the expected $T=0$ SG transition.
Another point to be noticed is that
$g_{{\rm SG}}$ for larger sizes exhibits a wavy structure
at around $0.02\lsim T/J\lsim 0.03$, suggesting the occurrence of
some sort of changeover in the ordering behavior.

\begin{figure}[h]
\includegraphics[width=\figwidth]{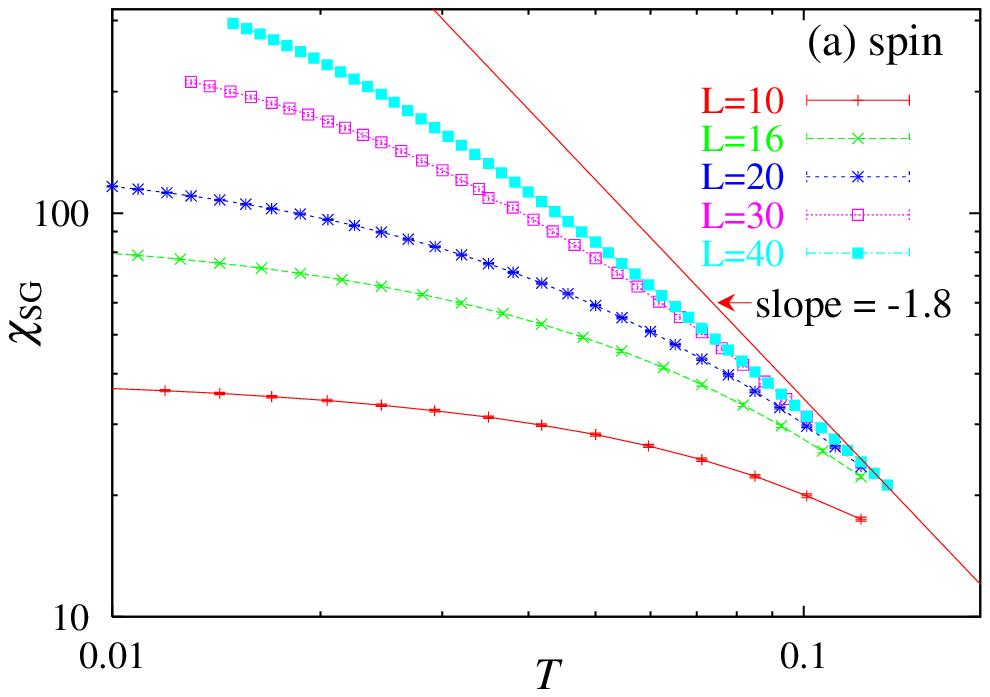}
\includegraphics[width=\figwidth]{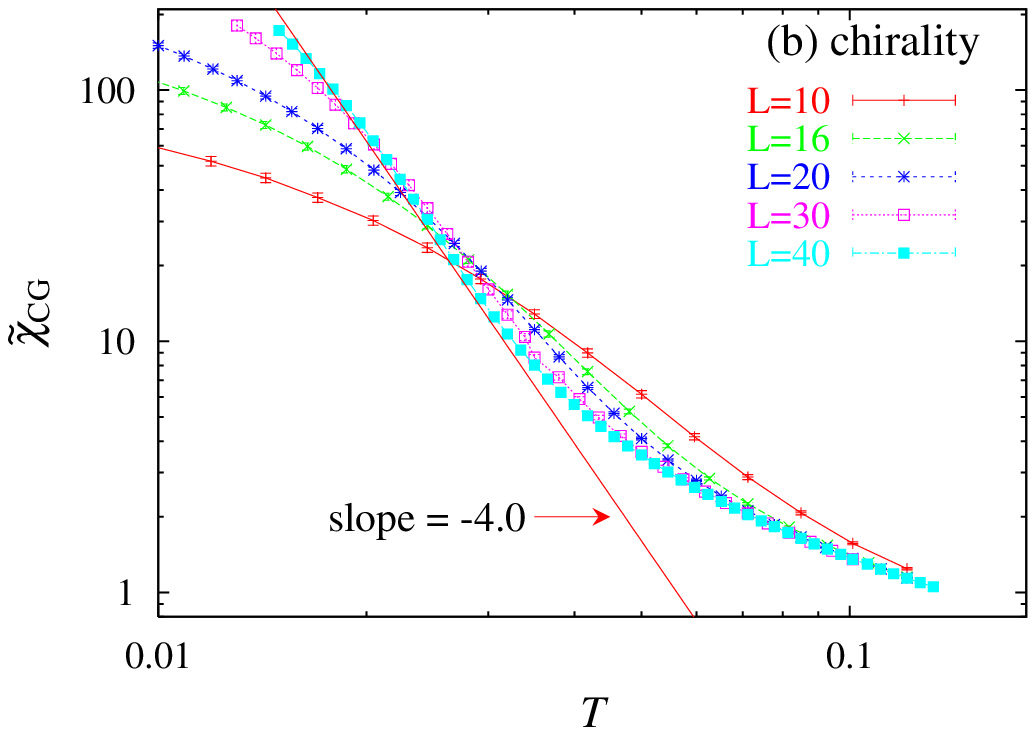}
\caption{
Temperature dependence of (a) the spin-glass susceptibility, and (b)
the reduced chiral-glass susceptibility (b), on a log-log plot.
}
\end{figure}

Similarly to the spin Binder ratio,
the chiral Binder ratio  for smaller sizes $L\leq 20$ also 
exhibits a crossing
at a nonezero temperature,
where the crossing
temperautres  tend to shift to lower temperatures with increasing $L$.
For larger sizes $L\geq 20$, by contrast, $g_{{\rm CG}}$
does not exhibit a crossing
at any temperature studied,
always getting smaller with increasing $L$. Such a behavior strongly
suggests that the CG transition 
occurs only at $T=0$.
A closer look of the data
reveals that $g_{{\rm CG}}$ exhibits a shallow negative
dip at a size-dependent temperature $T_{dip}(L)$, which tends to become
shallower with increasing $L$. Although such a
negative dip is also observed in the 3D Heisenberg SG, it tends to become
deeper in the 3D case in contrast to the present 2D case\cite{HK1}.
As argued in Ref.\cite{HK1},
if a negative dip persists 
in the $L\rightarrow \infty$ limit with  $T_{dip}(L=\infty)>0$, it means
the occurrence of a finite-temperature CG transition at
$T=T_{{\rm CG}}=T_{dip}(L=\infty)$. In the present 2D case, however,
the observed $T_{dip}(L)$ decreases rapidly with increasing $L$,
tending to $T=0$ in the $L\rightarrow \infty$ limit. Indeed,
a fit of our data of $T_{dip}(L)$
to the form,  $T_{dip}(\infty)+{\rm const.}/L$, yields an estimate
$T_{dip}(\infty)/J=0.00\pm 0.01$. 
Thus, we conclude again that the CG transition
of the present 2D model occurs only at $T=0$,
simultanesouly with the SG transition.

\begin{figure}[h]
\includegraphics[width=\figwidth]{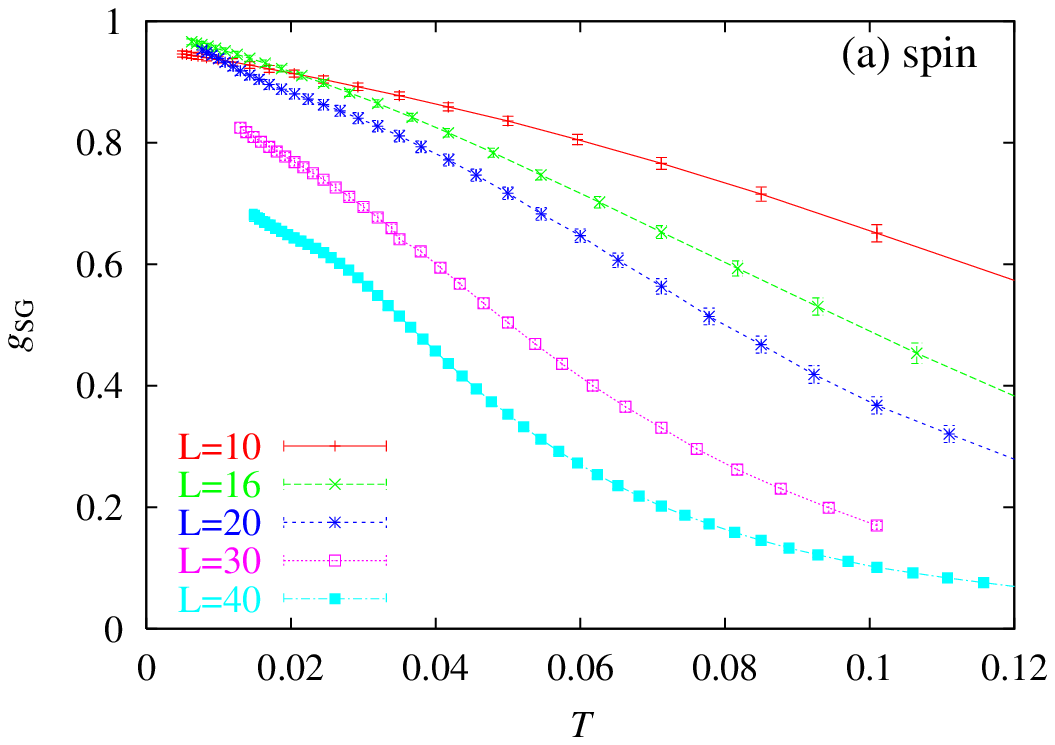}
\includegraphics[width=\figwidth]{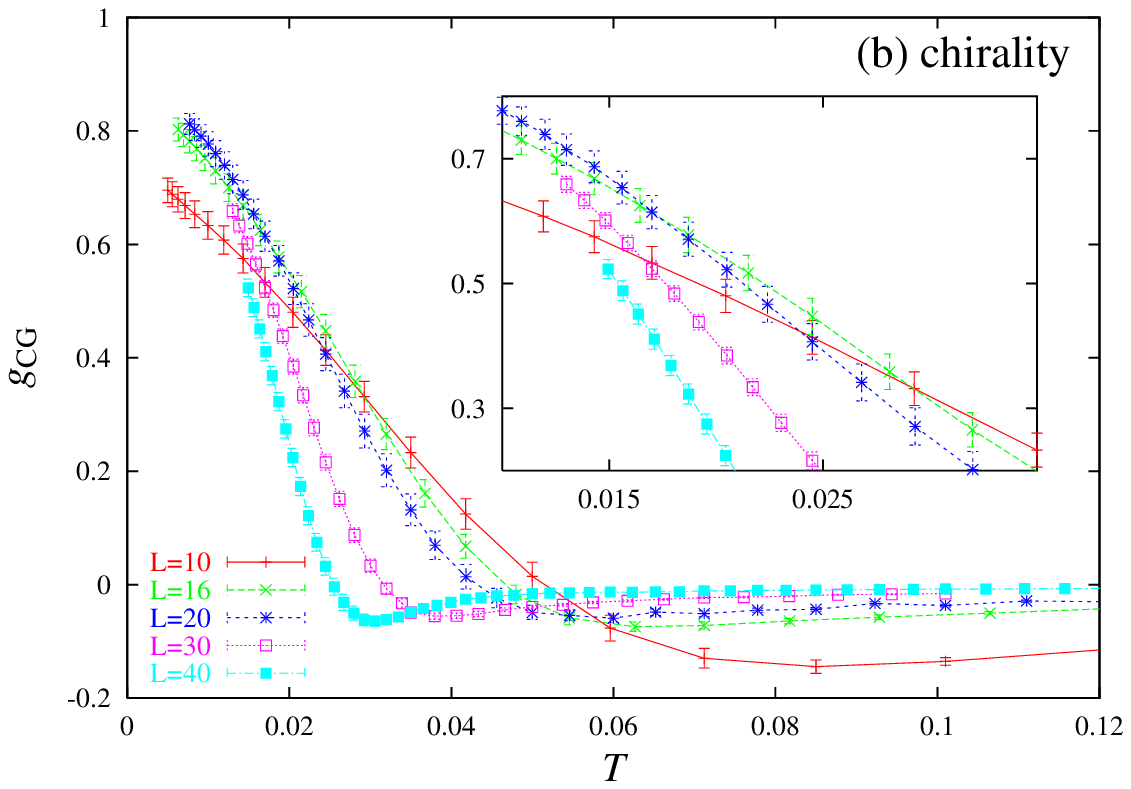}
\caption{
Temperature dependence of (a) the spin Binder ratio, and (b)
the chiral Binder ratio. The inset is a magnified view of the
low-temperature region.
}
\end{figure}

Next, we show our data of the correlation lengths.
Fig.3 exhibits the temperature and size dependence of the SG
correlation length $\xi_{{\rm SG}}$.
One sees from the figure that $\xi_{{\rm SG}}$
exhibits a diverging behavior toward the zero-temperature
transition point. We note that the size dependence of
$\xi_{{\rm SG}}$ is normal, always getting larger with increasing
$L$.

Figs. 4(a)-(c) exhibit the temperature and size dependence of the three
distinct types of the CG correlation lengths, $\xi_{{\rm CG}}^\perp$,
$\xi_{{\rm CG}}^\parallel$ and $\xi_{{\rm CG}}^+$, defined
above. As can be seen from the
figures, these three chirality correlation lengths exhibit
very much similar behaviors.
They all show a diverging behavior toward the $T=0$ 
transition point, accompanied by similar
size dependence. While they all show  normal size dependence at lower
temperatures $T/J\lsim 0.03$, increasing with  $L$,
they all show unusual size dependence
in the range $T/J\gsim 0.03$, decreasing with $L$.

\begin{figure}[h]
\includegraphics[width=\figwidth]{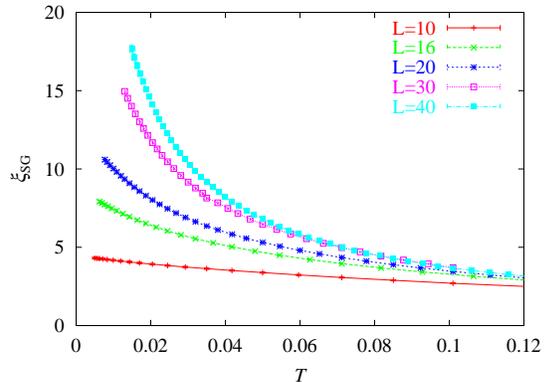}
\caption{
Temperature dependence of the spin-glass correlation length.
}
\end{figure}

In order to compare the SG and the CG correlation lengths more
quantitatively, we plot in Fig.5 $\xi_{{\rm SG}}$
together with $\xi_{{\rm CG}}^+$, which is the longest one among the
three types of CG correlation lengths.
At higher temperatures, the chiral correlation
length is shorter than the SG correlation length. This simply reflects
the fact  that the  short-range correlation of the spins is a
prerequisite for the onset of any chiral order.
As the temperature is decreased, $\xi_{{\rm CG}}$ grows much faster than
$\xi_{{\rm SG}}$, then catches up and eventually exceeds $\xi_{{\rm SG}}$
at a size-dependent
temperature $T_\times (L)$. This behavior suggests
that the chirality orders more
strongly than the spin. As shown in Fig.5,
the linear fit of the log-log plot of the data 
yields  slopes $\nu_{{\rm SG}}=0.9\pm 0.2$ for $\xi_{{\rm SG}}$, and
$\nu_{{\rm CG}}=2.2\pm 0.3$ for $\xi_{{\rm CG}}$,
the latter being more than 
twice the former. These values are consistent with
our estimates of the SG and CG susceptibility exponents given above,
$\gamma_{{\rm SG}}=2\nu_{{\rm SG}}\simeq 1.8$ 
and $\gamma_{{\rm CG}}=2\nu_{{\rm CG}}\simeq 4.0$.

We note that
the crossover temperatures $T_\times (L)$ where $\xi_{{\rm CG}}$ exceeds
$\xi_{{\rm SG}}$ tend to decrease with increasing $L$. We get
$T_\times/J \simeq 0.025, 0.019, 0.017, 0.012$ and 0.012
for $L=10$, 16, 20, 30 and 40, respectively (for $L=30$ and 40,
we have extrapolated 
the data slightly to lower temperatures to 
estimate $T_\times$). In particular,
if $T_\times (L)$
would tend to zero as $L\rightarrow \infty$,
the spin correlation would dominate
the chiral correlation at any nonzero 
temperature. This seems not to be
the case here, however, 
since the estimated $T_\times$ for our two largest sizes
$L=30$ and 40 turn out to be almost the same,  $T_\times/J\simeq 0.012$.
Hence, the indication is that, even in an infinite system, $\xi_{{\rm CG}}$
exceeds $\xi_{{\rm SG}}$ at low enough  temperatures
$T<T_\times \sim 0.01J$, 
where the chiral correlation dominates the spin correlation.
The nontrivial question still to be addressed is
the asymptotic critical behaviors of
$\xi_{{\rm SG}}$ and of $\xi_{{\rm CG}}$ 
realized below $T_\times$. Although it is
difficult to say something definite in this regime because of the
lack of the data, we will discuss this point later.

\begin{figure}[h]
\includegraphics[width=\figwidth]{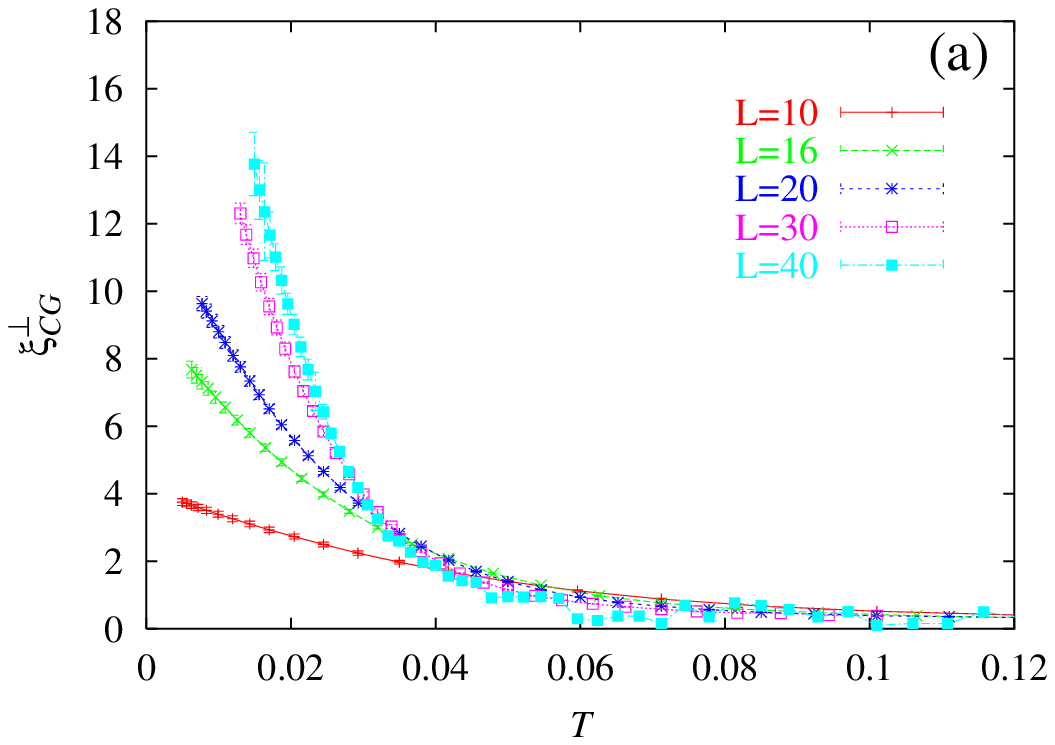}
\includegraphics[width=\figwidth]{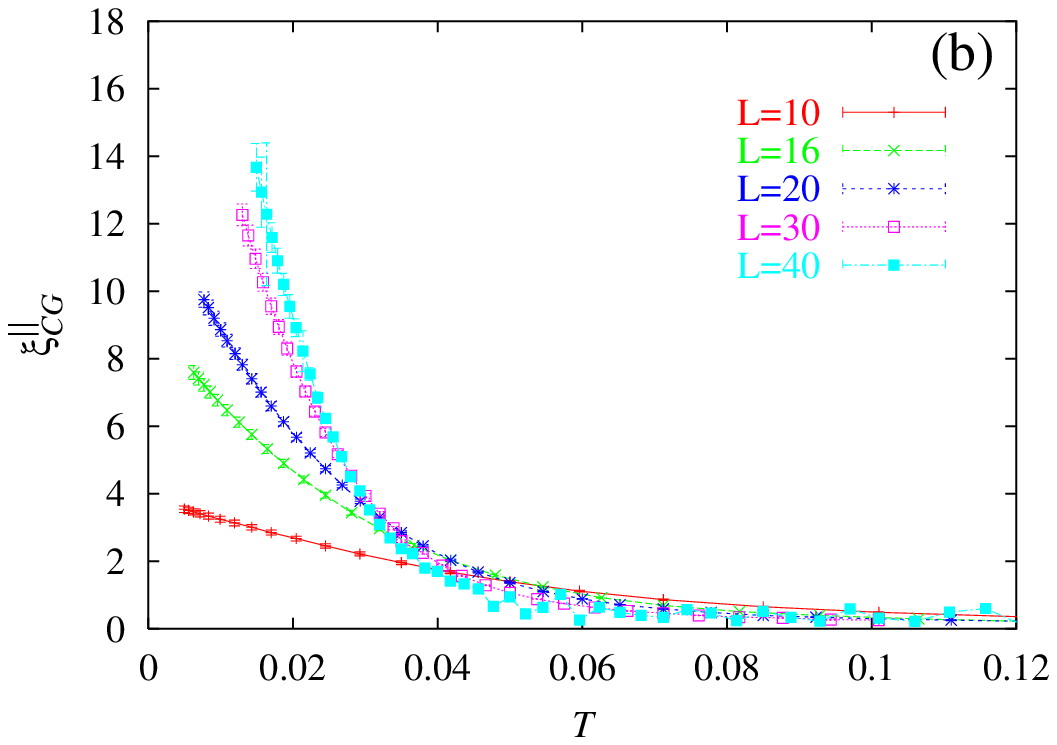}
\includegraphics[width=\figwidth]{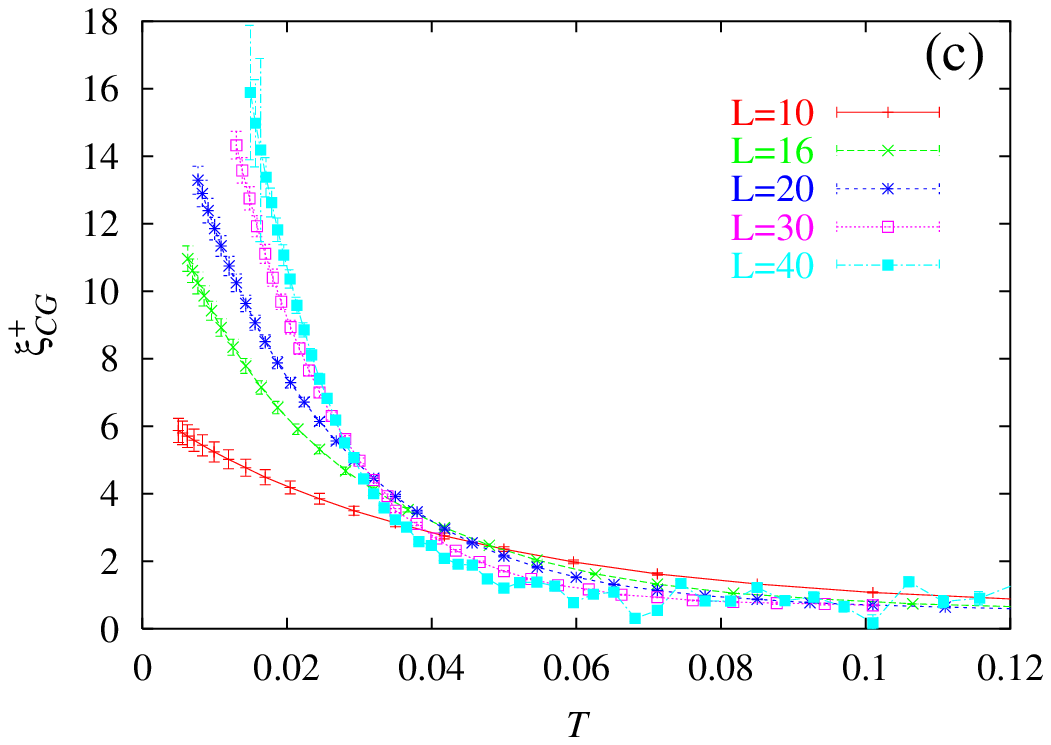}
\caption{
Temperature dependence of the three types of the chiral-glass correlation 
lengths, Eqs.(17)-(19); (a) $\xi_{{\rm CG}}^\perp$, (b)
$\xi_{{\rm CG}}^\parallel$, and (c) $\xi_{{\rm CG}}^+$. 
}
\end{figure}

In Fig.6(a), we show the temperature dependence of the
dimensionless SG correlation length $\xi_{{\rm SG}}/L$ for several sizes.
Somewhat unexpectedly, $\xi_{{\rm SG}}/L$
for smaller sizes $L=10,16$ and 20 cross
almost at a point $T/J\simeq 0.022$, giving an appearance of 
a finite-temperature SG transition.  However,
$\xi_{{\rm SG}}/L$ for larger sizes $L=30$ and 40 changes its behavior: They
gradually come down, without a crossing at $T/J=0.022$.
Hence, the behavior of $\xi_{{\rm SG}}/L$ for larger $L$ 
is eventually consistent with the SG transition only at $T=0$. However,
we note that this
behavior becomes evident only for larger sizes $L\gsim 20$, and one
has to be  careful in the interpretation of the $\xi_{{\rm SG}}/L$ data
for smaller lattices. If one looked at the present $\xi_{{\rm SG}}/L$ data
for smaller sizes only, say, for $L\leq 20$, one might 
easily misconclude that
the SG transtiion of the model
occurred at a nonzero temperature $T/J\simeq 0.022$.

\begin{figure}[h]
\includegraphics[width=\figwidth]{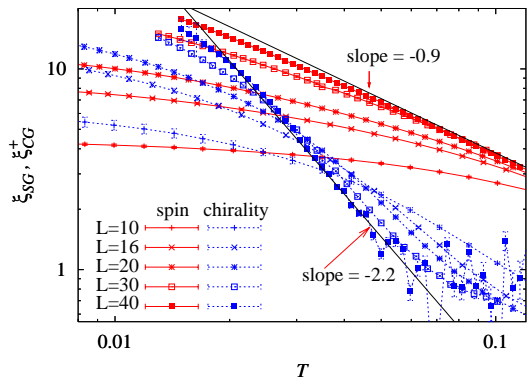}
\caption{
Temperature dependence of the spin-glass correlation length $\xi_{{\rm SG}}$
as compared with the chiral-glass correlation length $\xi_{{\rm CG}}^+$
on a log-log plot.
}
\end{figure}

In Fig.6(b), we show the temperature dependence of the
dimensionless CG correlation length $\xi_{{\rm CG}}/L$ for several sizes.
Again, $\xi_{{\rm SG}}/L$ for smaller size exhibits a crossing at a 
nonzero temperature, while
the crossing temperatures tend to shift down to lower temperatures in this
case. The observed behavior is consistent with the
occurrence of a $T=0$ CG transition. We note that 
very much similar
behaviors are found for the other types of the dimensionless CG correlation 
length, 
$\xi_{{\rm CG}}^\perp/L$ and $\xi_{{\rm CG}}^\parallel/L$
(data not shown here).

Having established that both the SG and the CG transitions
occur only at $T=0$, we now
analyze the critical properties associated with the $T=0$ transition
on the basis of a finite-size scaling. In the analysis below,
we use the data only at lower temperatures $T/J\leq 0.024$ where the
chirality enters into the critical regime 
exhibiting the normal size dependence,
and for moderately
large sizes $L\geq 20$. Since the exponent $\eta$ associated with
the $T=0$ transition
should be zero, we fix $T_{{\rm SG}}=0$, $T_{{\rm CG}}=0$,
$\eta_{{\rm SG}}=0$ and $\eta_{{\rm CG}}=0$ below.

We begin with the chirality ordering.
Fig.7(a) exhibits a finite-size scaling plot of
the CG suceptibility,
where the best value of $\nu_{{\rm CG}}$ is estimated
to be $\nu_{{\rm CG}}=2.0\pm 0.3$. 
The value of $\nu_{{\rm CG}}$ is consistent with
the one determined  from the log-log plot of the CG
susceptibility.

In Fig.7(b), we show a scaling plot of the chiral Binder ratio $g_{{\rm CG}}$
with $\nu_{{\rm CG}}=2.0$. The data collapse turns out to be
reasonably good.
In Fig.7(c), we show a scaling plot of the dimensionless
chiral correlation length $\xi_{{\rm CG}}/L$
with $\nu_{{\rm CG}}=2.0$. The data collapse turns out
to be marginally good.
Hence, our data of the CG susceptibility, the chiral Binder ratio
and the chiral correlation length seem all consistent with 
$\nu_{{\rm CG}}=2.0\pm 0.3$. Taking account of the corresponding
estimate based on Fig.5, we finally quote $\nu_{{\rm CG}}=2.1\pm 0.3$
as our best estimate of $\nu_{{\rm CG}}$.

\begin{figure}[h]
\includegraphics[width=\figwidth]{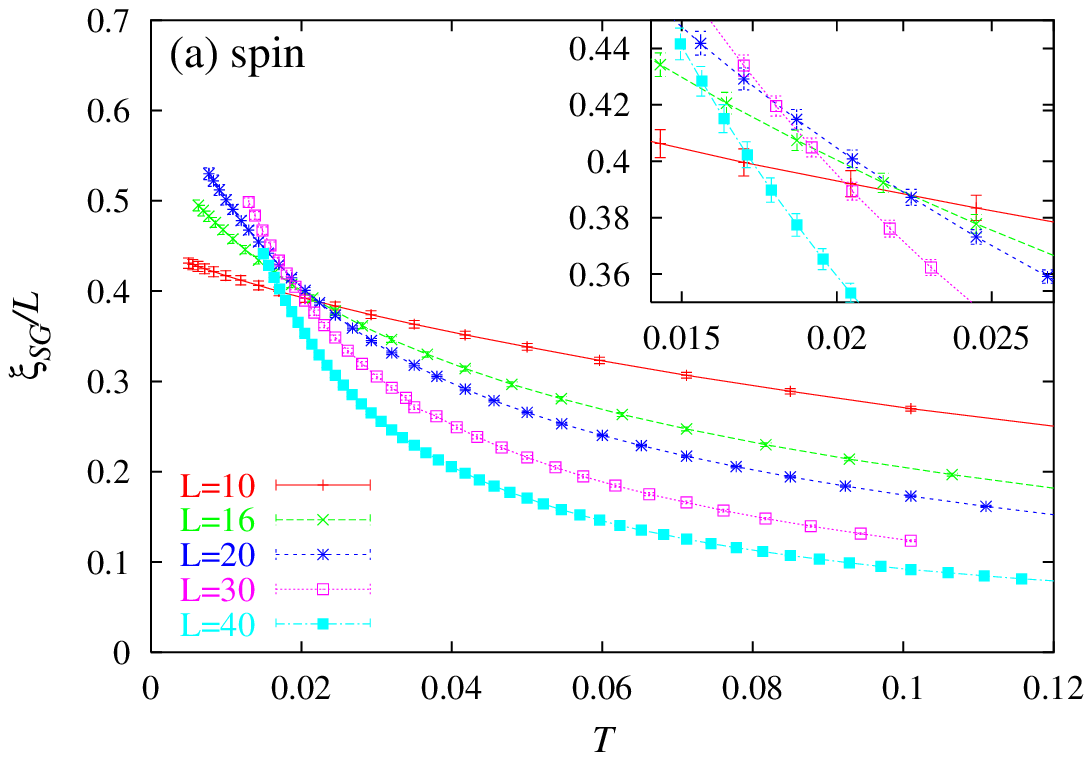}
\includegraphics[width=\figwidth]{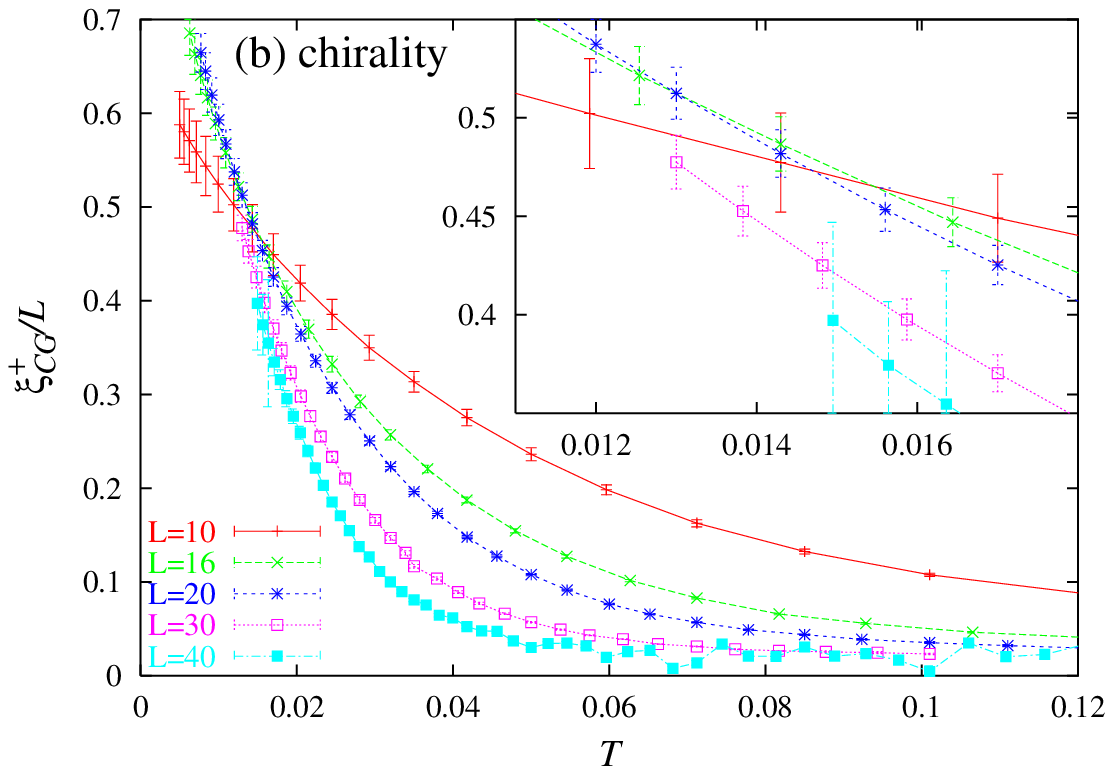}
\caption{
Temperature and size dependence of (a) the dimensionless 
spin-glass correlation length $\xi_{{\rm SG}}/L$, and (b)
the dimensionless 
chral-glass correlation length $\xi_{{\rm CG}}^+/L$.
The insets are magnified views of the
low-temperature region.
}
\end{figure}

Next, we examine the spin ordering.
Fig.8(a) exhibits a finite-size scaling plot of
the SG suceptibility,
where the best value of $\nu_{{\rm SG}}$ is estimated
to be $\nu_{{\rm SG}}=0.9\pm 0.2$. 
The value of $\nu_{{\rm SG}}$ is consistent with
the one determined above from the log-log plot of the SG susceptibility.
In Fig.8(b), we show a scaling plot of the spin Binder ratio $g_{{\rm SG}}$
with $\nu_{{\rm SG}}=0.9$.
The data collapse is rather poor here, however.
The wavy structure of the $g_{{\rm SG}}$
curve observed for larger $L$ turns out to deteriorate 
the data collaplse. The quality of
the plot does not improve even with other choices of $\nu_{{\rm SG}}$.
In Fig.8(c), we show a scaling plot of the dimensionless
SG correlation length $\xi_{{\rm SG}}/L$
with $\nu_{{\rm SG}}=0.9$. Again, the data collapse is poor, and
does not improve even with other choices of $\nu_{{\rm SG}}$.

\begin{figure}[h]
\includegraphics[width=\figwidth]{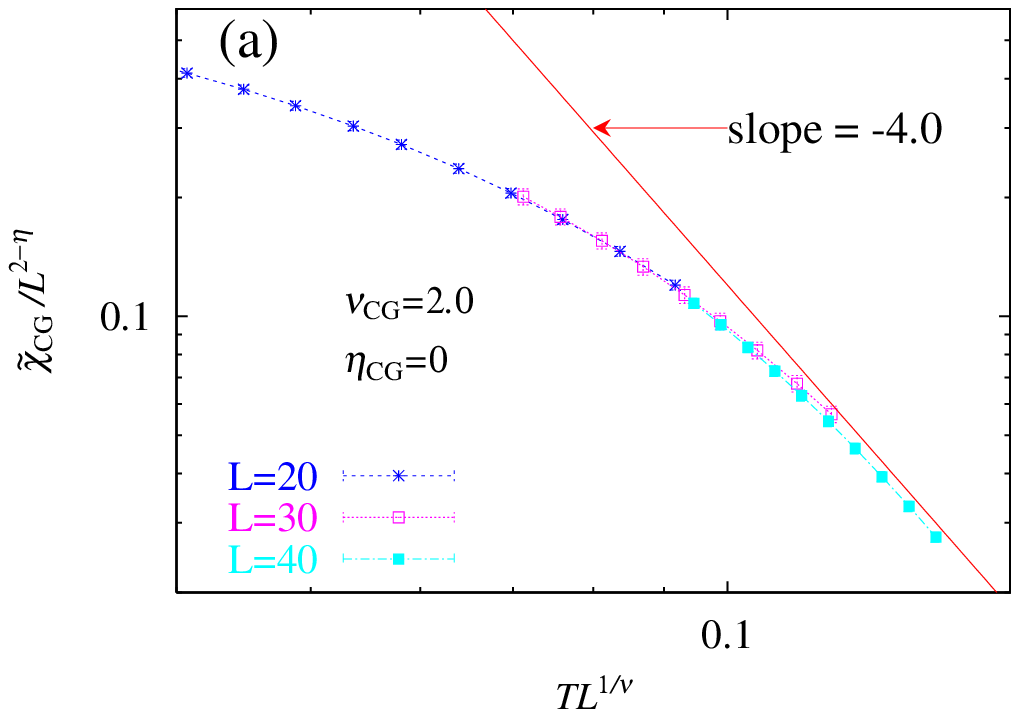}
\includegraphics[width=\figwidth]{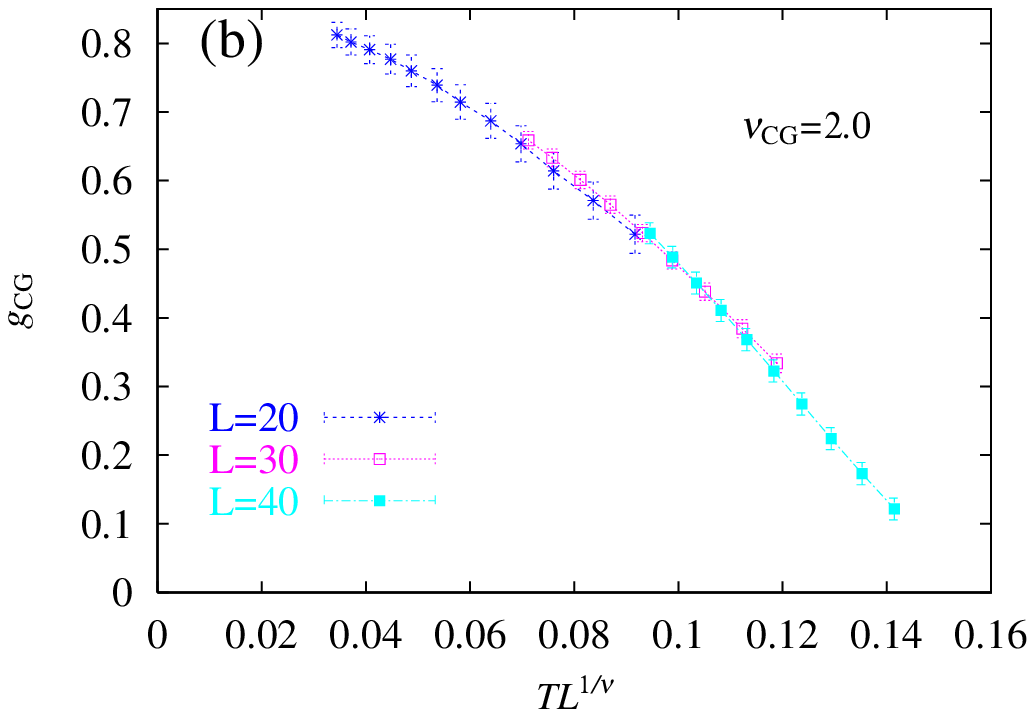}
\includegraphics[width=\figwidth]{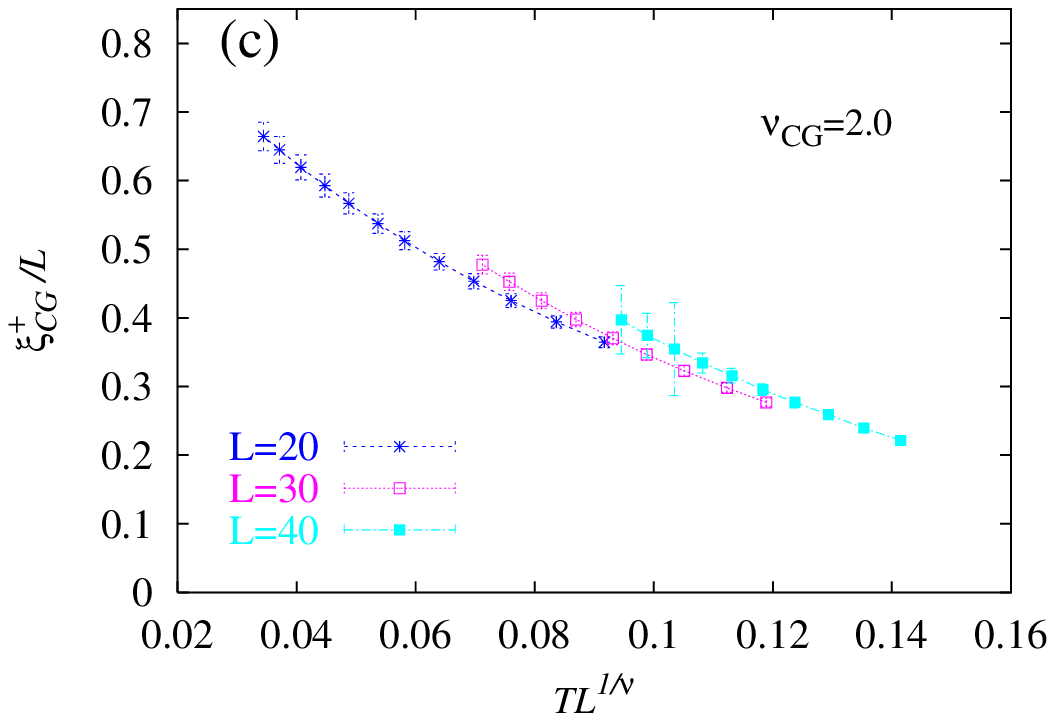}
\caption{
Finite-size scaling plots of (a) the chiral-glass susceptibility,
(b) the chiral Binder ratio, and (c) the dimensionless chiral-glass 
correlation length.
The chiral-glass transition temperature is $T_{{\rm CG}}=0$.
}
\end{figure}

Hence, although
the SG susceptibility can be
scaled by assuming $\nu_{{\rm SG}}\simeq 0.9$, neither
the spin Binder ratio nor the SG correlation length
can be scaled by the same $\nu_{{\rm SG}}$, nor by assuming any other 
$\nu_{{\rm SG}}$.
Possible reason for this is either of the followings.

\begin{figure}[h]
\includegraphics[width=\figwidth]{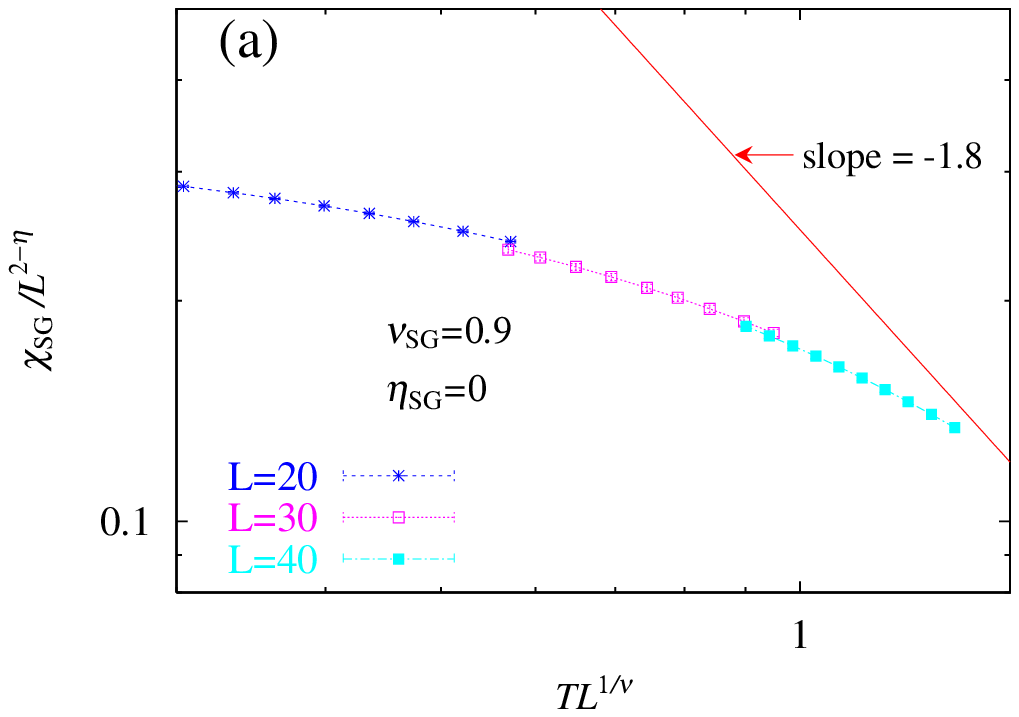}
\includegraphics[width=\figwidth]{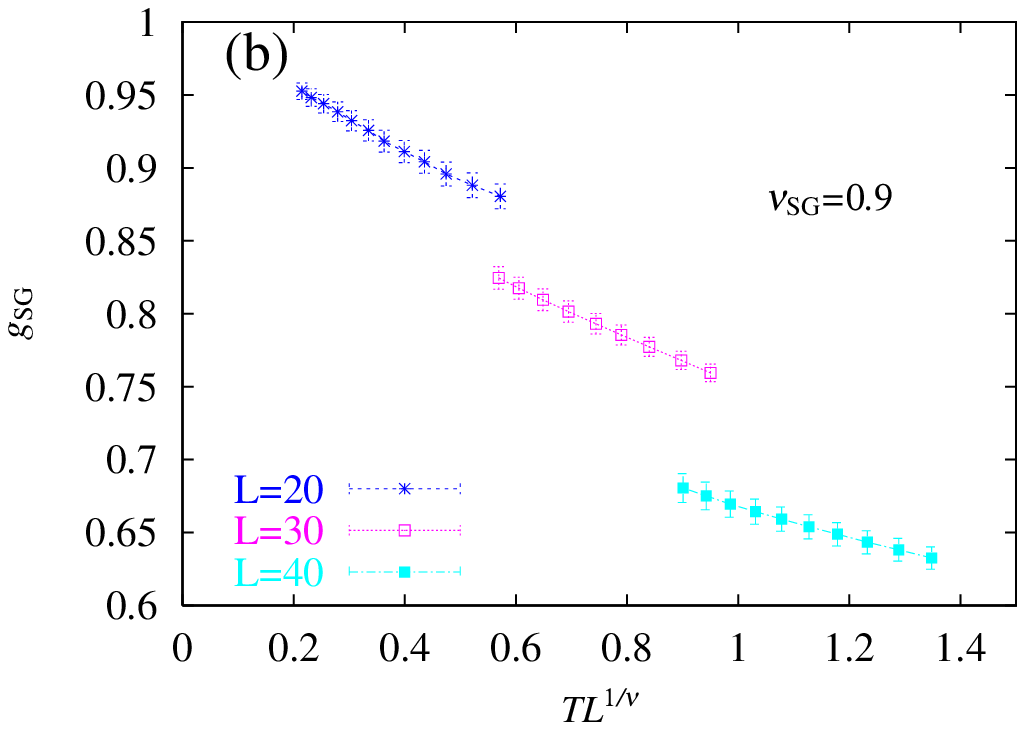}
\includegraphics[width=\figwidth]{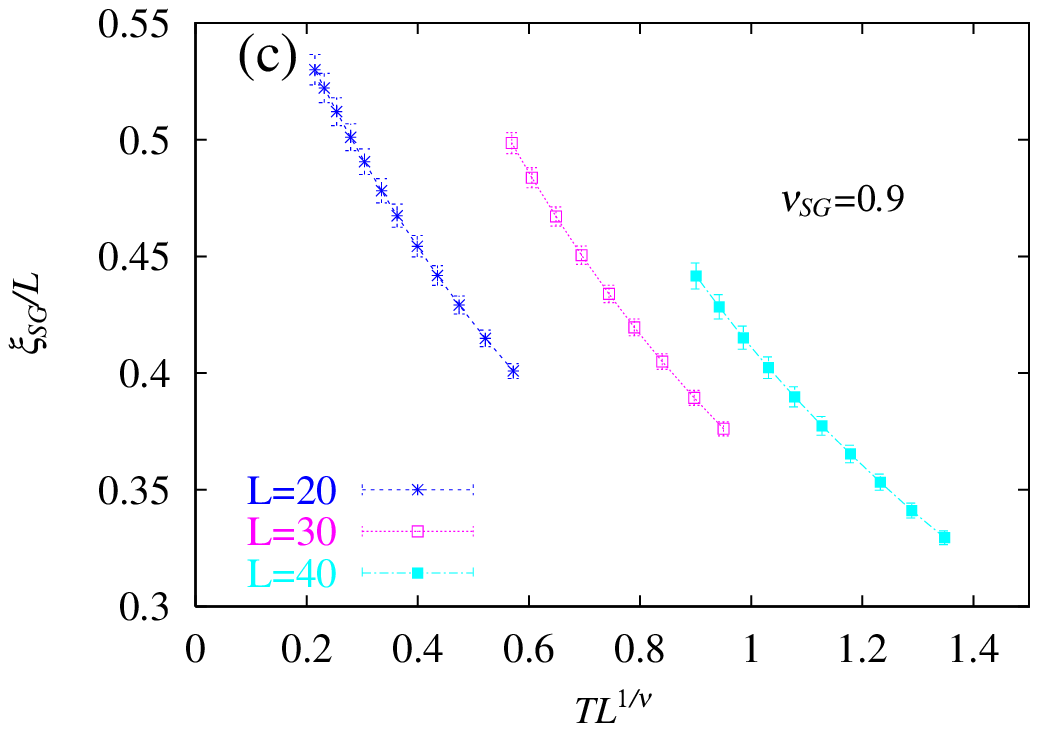}
\caption{
Finite-size scaling plots of (a) the spin-glass susceptibility,
(b) the spin Binder ratio, and (c) the dimensionless spin-glass
correlation length.
The transition temperature is  $T_{{\rm SG}}=0$.
}
\end{figure}

i) The width of the critical regime associated with the $T=0$ SG transition
depends on the physical quantities measured, being wide for the SG
susceptibility but narrower for the Binder ratio or the correlation length.
In this case, $\nu_{{\rm SG}}\simeq 0.9$ determined above from 
$\chi_{{\rm SG}}$ is a true asymptotic exponent, and
at low enough temperatures and for large enough sizes, both $g_{{\rm SG}}$
and $\xi_{{\rm SG}}$ should evnetually scale with $\nu_{{\rm SG}}
\simeq 0.9$.

ii) The scaling behavior observed in $\chi_{{\rm SG}}$ is
somewhat accidental, not associated with the true asymptotic one.
At low enough temperatures
where $\xi_{{\rm CG}}$ well exceeds $\xi_{{\rm SG}}$, the scaling behavior
of $\chi_{{\rm SG}}$ is changed into a different one characterized by
a different $\nu_{{\rm SG}}$ value, 
which would also scale both $g_{{\rm SG}}$
and $\xi_{{\rm SG}}$ at low enough temperatures and for large enough
sizes. Unfortunately,
the lack of our data in the low temperature regime $T\lsim T_\times$ prevents
us from giving any sensible estimate of the asympotic $\nu_{{\rm SG}}$ value.
If the latter possibility ii) is really the case, 
there even exists a possibility that the
spin-chirality decoupling eventually does not occur so that one has
$\nu_{{\rm SG}}=\nu_{{\rm CG}}$. If one trusts our estimate of $\nu_{{\rm CG}}$
given above, $\nu_{{\rm CG}}\simeq 2.1$, the absence of the
spin-chirality decoupling would then mean $\nu_{{\rm SG}}\simeq 2.1$.
Although we cannot completely
rule out such a possibility, we also mention that
any sign of such an equality $\nu_{{\rm SG}}=\nu_{{\rm CG}}$ was not seen
in the available data.

At present, we cannot tell for sure which of the above possibilities,
i) or ii), is really the case. 
Nevertheless, in view of the
scaling observed in $\chi_{{\rm SG}}$, 
it seems to us that the above scenario ii) is rather unlikely, 
and the scenario i) is more plausible. Then, 
true asymptotic spin and chirality correlation-length exponents
would not be far from
$\nu_{{\rm SG}}=0.9\pm 0.2$ and $\nu_{{\rm CG}}=2.1\pm 0.3$
quoted above, which
means the occurrence of the spin-chirality decoupling in the
2D Heisenberg SG. We note that these exponent values are close  to 
the values reported in Ref.\cite{Kawamura92} for the same model
by means of the $T=0$ numerical 
domain-wall-energy calculation, 
$\nu_{{\rm SG}}= 1.0\pm 0.15$ and $\nu_{{\rm CG}}= 2.1\pm 0.2$.

%
%
\section{Summary}

In summary, we performed large-scale equilibrium
MC simulations of the 2D Heisenberg spin glass.
Particular attention was paid to the behavior of the spin and the chirality
correlation lengths.
In order to observe the true
asymptotic behavior, fairly large system size $L\gsim 20$ appears to
be necessary.
We have established
that both the spin and the chirality order only at $T=0$.
The model seems to possess the two characteristic temperatures 
scales $T^*$ and
$T_\times$ ($T^*>T_\times >0$). Around $T^*$, the model enters into
the chiral critical regime where critical fluctuations of the chirality
become significant. This shows up as a changeover in
the size dependence of the
chiral correlation length or the CG susceptibility: Namely, above $T^*$,
$\xi_{{\rm CG}}$ and  $\tilde \chi_{{\rm CG}}$ tend to decrease with $L$, 
whereas below $T^*$ they tend to increase with $L$. 
The non-critical
behavior observed above $T^*$ might be understood as a
mean-field-like behavior, as argued 
in Ref.\cite{ImaKawa}. 
The chiral correlation length, which stays shorter than the spin correlation
length at high temperatures, cathes up and
exceeds the spin correlation length around the second characteristic
temperature, $T_\times\simeq 0.01J$. Our data in the temperature regime
$T_\times\lsim T\lsim T^*$ and for size $20\lsim L\lsim 40$ 
yields the estimates
of the spin and the chirality correlation-length
exponents, $\nu_{{\rm SG}}=0.9 \pm 0.2$, $\nu_{{\rm CG}}=2.1 \pm 0.3$,
respectively. Unfortunately, we donot have much data in the
interesting temperature range
$T\lsim T_\times$, which prevents us from directly determining the
truly asymptotic critical
properties associated with the $T=0$ transition. While we speculate that
the asymptotic spin and the chirality exponents 
would not be far from the values quoted above,
the data in the lower temperature regime $T\lsim T_\times \simeq 
0.01J$ and for larger sizes $L\gsim 40$ is required
to settle the issue.

%
\section*{Acknowledgements}

The numerical calculation was performed on the HITACHI SR8000
at the supercomputer system, ISSP, University of Tokyo.
The authors are thankful to Dr. H. Yoshino and K. Hukushima for
useful discussion.



\end{document}